\documentclass[iop]{emulateapj}
\usepackage{apjfonts}
\usepackage[breaklinks,colorlinks,urlcolor=blue,citecolor=blue,linkcolor=blue]{hyperref}
\journalinfo{The Astrophysical Journal, 2015, in press}

\shorttitle{Driving Spicules and Jets with MHD Turbulence}
\shortauthors{Cranmer \& Woolsey}

\begin{document}

\title{Driving Solar Spicules and Jets with Magnetohydrodynamic
Turbulence: Testing a Persistent Idea}

\author{Steven R. Cranmer\altaffilmark{1}
and Lauren N. Woolsey\altaffilmark{2}}

\altaffiltext{1}{Department of Astrophysical and Planetary Sciences,
Laboratory for Atmospheric and Space Physics,
University of Colorado, Boulder, CO 80309, USA}
\altaffiltext{2}{Harvard-Smithsonian Center for Astrophysics,
60 Garden Street, Cambridge, MA 02138, USA}

\begin{abstract}
The solar chromosphere contains thin, highly dynamic strands of plasma
known as spicules.
Recently, it has been suggested that the smallest and fastest (Type II)
spicules are identical to intermittent jets observed by the
{\em Interface Region Imaging Spectrograph.}
These jets appear to expand out along open magnetic field lines rooted
in unipolar network regions of coronal holes.
In this paper we revisit a thirty-year-old idea that spicules may be
caused by upward forces associated with Alfv\'{e}n waves.
These forces involve the conversion of transverse Alfv\'{e}n waves
into compressive acoustic-like waves that steepen into shocks.
The repeated buffeting due to upward shock propagation
causes nonthermal expansion of the chromosphere and a transient
levitation of the transition region.
Some older models of wave-driven spicules assumed sinusoidal wave
inputs, but the solar atmosphere is highly turbulent and stochastic.
Thus, we model this process using the output of a time-dependent
simulation of reduced magnetohydrodynamic turbulence.
The resulting mode-converted compressive waves are strongly variable
in time, with a higher transition region occurring when the amplitudes
are large and a lower transition region when the amplitudes are small.
In this picture, the transition region bobs up and down by several Mm
on timescales less than a minute.
These motions produce narrow, intermittent extensions of the chromosphere
that have similar properties as the observed jets and Type~II spicules.
\end{abstract}


\section{Introduction}
\label{sec:intro}

The Sun's hot corona expands into interplanetary space as a supersonic
plasma outflow known as the solar wind.
However, we still do not yet know how the tenuous corona/wind system
is formed from the much larger pool of mass and energy in the
colder photosphere and chromosphere.
High-resolution observations \citep[e.g.,][]{Fl15}
show that coronal heating is highly dynamic and intermittent in
space and time.
Some have suggested that much of the corona's mass and energy may
be injected in the form of narrow features known variously as
spicules, fibrils, and mottles \citep{Be72,St00,Dp07,Ts12}.
Recently, the {\em Interface Region Imaging Spectrograph}
({\em IRIS,} \citeauthor{Dp14a}~\citeyear{Dp14a})
found similar jet-like features emerging
from largely unipolar network flux concentrations in coronal holes
and quiet regions on the solar surface \citep{Ti14,Ti15}.

Coronal holes have long been known to play host to narrow, ray-like
features known as polar plumes and polar jets
\citep{NH68,AW77,Wa98,Do02,Cu07,Ra08,YM14,Pv15}.
These bright strands typically extend up to heights of order
0.1--1~$R_{\odot}$, whereas spicules and the {\em IRIS} network jets
have length scales of only $\sim$0.01~$R_{\odot}$.
The largest plumes and jets are often associated with the emergence of
small magnetic loops at their footpoints, and thus they are believed
to be powered by magnetic reconnection
\citep[see, e.g.,][]{WS95,Sh07,Pa09,Mo10,Cg15}.
However, no clear evidence for reconnection in the smaller
{\em IRIS} jets has been found so far.

In this paper we explore the idea that magnetohydrodynamic (MHD) waves
are responsible for producing rapidly varying, field-aligned extensions
of cool chromospheric gas along network flux tubes in coronal holes.
These modeled features are found to have similar properties as
the {\em IRIS} network jets, which \citet{Ti14} proposed to be
identical to the so-called Type~II spicules observed above the limb.
The idea that spicules could be driven by MHD waves has been discussed
for several decades \citep{HJG,MH85,Ho92,KS99,Dp99,VL99,MS10,Mu15},
but it is still not known whether this is a dominant mechanism in
the real solar atmosphere.

Many of the necessary ingredients of the wave-driven spicule model
are supported by observations.
The solar corona contains transverse, incompressible oscillations in
the magnetic field and plasma velocity \citep{Bj98,To07,Js09,Mc12},
but whether they should be called ``Alfv\'{e}n waves'' is still a
matter of debate \citep[e.g.,][]{Ma13}.
More specifically, spicules and jets themselves seem to contain
torsional or kink motions that could have an Alfv\'{e}nic character
\citep{Ku06,Dp14b,Tv15}.
There is also evidence for longitudinal compressive waves---i.e.,
fluctuations in density and the velocity component parallel to the
wavenumber vector---that may or may not follow the ideal MHD
magnetoacoustic dispersion relations \citep{Of99,KP12,Th13,My14}.
In polar plumes, \citet{Li15} found high-frequency Alfv\'{e}n-like
waves and low-frequency compressive waves traveling along the same
field lines, which points to the possibility of mode coupling.
Also, \citet{Pa15} found what appear to be slow-mode magnetosonic
waves within the {\em IRIS} network jets.

The models developed in this paper rely on MHD waves behaving in an
intermittent and stochastic manner.
Several earlier attempts to understand spicules as a by-product of
waves assumed a periodic, sinusoidal driver at the lower boundary of
the modeled system.
However, both observations \citep{TM09,Li14} and simulations
\citep[e.g.,][]{vB11,PC13,Zd15} show that MHD fluctuations in the
chromosphere and corona exhibit continuous power-law spectra and
irregular bursts of activity and dissipation.
This variability may also be related to the fact that the Sun's
transition region (TR) has a complex ``corrugated'' shape
\citep{Fe79,Zh98,Pe13}.
We propose that jets and Type~II spicules are short-lived extensions
of the corrugated TR that are driven by similarly infrequent outliers
in the underlying population of waves and turbulent eddies.

The remainder of this paper is organized as follows.
Section \ref{sec:mode} discusses how chromospheric Alfv\'{e}n waves
may evolve nonlinearly into a collection of compressible
fluctuations.
In Section \ref{sec:steep} we estimate the degree of nonthermal
expansion experienced by the upper chromosphere and TR as a result of
shocks that develop from a compressive wave train.
Section \ref{sec:braid} takes the output from a reduced MHD simulation
of Alfv\'{e}nic turbulence in a coronal hole and computes the
time-dependent generation of compressive waves and intermittent
levitation of the TR.
In Section \ref{sec:interp} we compare the modeled up-and-down
motions of the TR with the observed properties of Type~II spicules
and {\em IRIS} network jets.
Lastly, in Section \ref{sec:conc} we summarize our results, discuss some
broader implications, and describe future improvements to the models.

\section{Nonlinear Production of Parallel Velocity Fluctuations}
\label{sec:mode}

When incompressible Alfv\'{e}nic fluctuations grow to a sufficiently
large amplitude (i.e., when the oscillating transverse magnetic field
$\delta B_{\perp}$ becomes of the same order of magnitude as the
background field strength $B_0$), they become susceptible to a range of
nonlinear interactions that can spawn other types of waves.  For example,
\begin{enumerate}
\item
It has been known for several decades that linearly polarized Alfv\'{e}n
waves can excite second-order {\em ponderomotive oscillations} in
density, gas pressure, and magnetic pressure
\citep{Ho71,Sp89,VH96}.
Because these oscillations are tied to the extrema of transverse
arcs traced by the magnetic field vector, their frequencies tend to
be twice those of the original Alfv\'{e}n wave.
Corresponding wave periods, for Alfv\'{e}n waves oscillating with
${\cal P}_{\rm A} \approx 3$--5~minutes, are of order 1--2 minutes.
These are reminiscent of the durations and recurrence timescales of
Type~II spicules.
\item
There has also been substantial work done to study the nonlinear
development of {\em parametric instabilities} for circularly
polarized Alfv\'{e}n waves \citep{Go78,JH93,TT03,DZ15}.
For conditions appropriate to the corona, this instability usually
involves an upward propagating Alfv\'{e}n wave decaying into a
downward propagating Alfv\'{e}n wave and an upward
propagating magnetosonic-like wave.
The latter tends to have a lower frequency $\omega_{\rm S}$
than that of the original Alfv\'{e}n wave $\omega_{\rm A}$.
Typical periods (${\cal P} = 2\pi/\omega$) in the open-field corona are
\begin{equation}
  {\cal P}_{\rm S} \, \approx \,
  \left( \frac{V_{\rm A}}{2 c_s} \right) {\cal P}_{\rm A}
\end{equation}
which for ${\cal P}_{\rm A} \approx 3$~minutes, typical coronal
Alfv\'{e}n speeds $V_{\rm A} \approx 2000$ km~s$^{-1}$, and sound speeds
$c_{s} \approx 150$ km~s$^{-1}$, gives a period of order 20 minutes,
similar to what is observed for density fluctuations above the limb
\citep[e.g.,][]{Of99,Th13}.
\end{enumerate}
The number of possible mode-coupling interactions grows even bigger
when the MHD waves pass through a strongly inhomogeneous background
medium \citep[e.g.,][]{HP83,LR86,Nk98,HK12,TM13}.

Because we are concerned with the development of short-lived spicules
and jets along unipolar field lines, we focus on the most rapid kind
of nonlinear mode coupling that occurs in homogeneous plasmas:
the second-order ponderomotive effect.
\citet{Ho71} showed that large-amplitude linearly polarized Alfv\'{e}n
waves produce an oscillation $\delta v_{\parallel}$ in the parallel
velocity with an amplitude that scales as
\begin{equation}
  \frac{\delta v_{\parallel}}{V_{\rm A}} \, = \, N_{\beta} \,
  \left( \frac{\delta B_{\perp}}{B_0} \right)^{2} \,\, ,
  \label{eq:convert}
\end{equation}
where $N_{\beta}$ describes the dependence on the plasma $\beta$
parameter.
We use a simplified kinetic definition of $\beta = (c_{s}/V_{\rm A})^2$
which is different by a factor of $2/\gamma \approx 1.2$ from
the standard MHD definition of the ratio of pressures.
\citet{Ho71} used the ideal MHD conservation equations to derive
\begin{equation}
  N_{\beta} \, = \, \frac{0.25}{|1 - \beta|} \,\, ,
\end{equation}
which diverges unrealistically at the asymptotic value of $\beta = 1$.
\citet{VH96} extended the second order MHD theory and found, for some
cases, $N_{\beta} = 0.25/(1 + \beta)$.
Several other properties of this solution were consistent with a
magnetosonic-like wave mode.

\begin{figure}
\epsscale{1.11}
\plotone{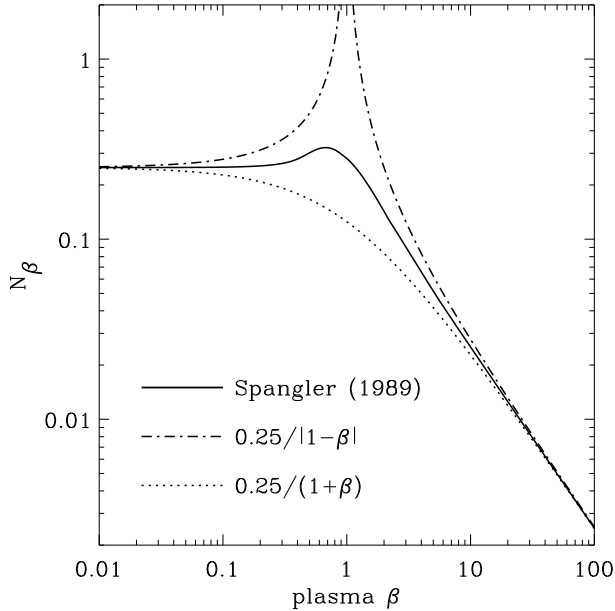}
\caption{Dependence of the dimensionless mode-conversion
efficiency $N_{\beta}$ on the kinetic plasma $\beta$, including
results from \citet{Ho71} (dot-dashed curve),
\citet{VH96} (dotted curve), and present work based on \citet{Sp89}
(solid curve).
\label{fig01}}
\end{figure}

A substantial amount of other theoretical work has been done to
simulate the nonlinear evolution of Alfv\'{e}n waves, much of which
involves solving the Derivative Nonlinear Schr\"{o}dinger (DNLS)
equation \citep[e.g.,][]{MW86,MD96}.
In this paper, we make use of the kinetic results of \citet{Sp89}, who
solved a perturbed Vlasov equation for the ponderomotive density
fluctuations associated with a soliton-like Alfv\'{e}nic pulse.
We solved the \citet{Sp89} equations for a range of plasma $\beta$
values and for the simple one-temperature case of $T_{e}=T_{p}$.
Effective compressional amplitudes were extracted from the simulated
density fluctuation profiles, which did not maintain the same
Lorentzian shape of the input Alfv\'{e}nic pulse.
We maintained continuity with earlier studies of sinusoidal waves by
computing the relative fluctuation ratio $\delta \rho / \rho_0$ 
as half the peak-to-peak pulse variation in density.
Also, we assumed the acoustic-like energy equipartition found by
\citet{Ho71},
\begin{equation}
  \frac{\delta v_{\parallel}}{V_{\rm A}} \, = \,
  \frac{\delta \rho}{\rho_0}
  \label{eq:equi}
\end{equation}
and used it to compute $\delta v_{\parallel}$ as in
Equation (\ref{eq:convert}).
The numerical results were used to construct the efficiency factor
$N_{\beta}$, and Figure~\ref{fig01} shows its dependence on $\beta$.
This function agrees with the above analytic results in the limits of
$\beta \ll 1$ and $\beta \gg 1$, and it falls in between them for
$\beta \approx 1$.
We fit the $\beta$ dependence of this factor by an approximate function,
\begin{equation}
  N_{\beta} \, \approx \,
  \frac{0.25}{\sqrt{1 + \beta^2}} +
  \frac{0.135 \, \beta^{2.4}}{0.305 + \beta^{4.6}}
  \label{eq:Nfit}
\end{equation}
and this is accurate to within about 3\% over the range of $\beta$
values shown in Figure~\ref{fig01}.
A plot of this function would be nearly indistinguishable from the
solid curve that shows the numerical results of the \citet{Sp89} model.

Later in this paper,
Equations (\ref{eq:convert}), (\ref{eq:equi}), and (\ref{eq:Nfit})
are used to compute the properties of acoustic-like waves that are
generated from a time-dependent simulation of Alfv\'{e}nic turbulence
in the chromosphere and low corona.
We find that there is a distinct local maximum in $\delta v_{\parallel}$
at a height $z_{b} = 876$~km in the low chromosphere.
That location will be taken as a conceptual ``base height'' and used
to compare against other models of acoustic wave evolution.

\section{Shock Steepening and Chromospheric Levitation}
\label{sec:steep}

In the solar chromosphere, upwardly propagating acoustic waves
undergo rapid growth in amplitude, with
$\delta v_{\parallel} \propto \rho_{0}^{-1/2}$ in the limit of an
isothermal atmosphere and no dissipation \citep{Lb08,Lb32}.
At some point, however, the amplitudes become large enough for the
waves to steepen into shocks \citep[e.g.,][]{SS72}
and thereby dissipate their energy into the surrounding atmosphere.
In the process of damping, the fluctuations are also able to exert
a mean upward {\em wave pressure gradient} force on the atmosphere
\citep{Dw70,J77}.
This force is essentially a net transfer of momentum from the
oscillations to the background gas.
In a hydrostatic atmosphere (or in the subsonic parts of the solar
wind), an increase in the total effective pressure increases the
gravitational scale height, and this in turn ``puffs up'' the cool
chromosphere.
This effect has also been explored in the context of atmospheres
of cool, evolved giant stars \citep[see, e.g.,][]{BC85,LK12}.

We simulated the above chain of events using a series of one-dimensional
time-steady solutions of the ZEPHYR code \citep{CvB07}.
This code produces a realistic description of the photosphere,
chromosphere, corona, and solar wind in open-field regions of the
solar atmosphere.
ZEPHYR solves equations of wave action conservation for both
acoustic and Alfv\'{e}n waves in the presence of several kinds of
dissipation and shock steepening, and it includes wave pressure terms
that couple the fluctuations to the background plasma.
In these models there is no coupling between the acoustic and Alfv\'{e}n
wave modes; they each evolve independently of one another.
Specifically, we made use of the grid of models from Section~8.2 of
\citet{CvB07}, in which the Alfv\'{e}n wave properties were held fixed
and the acoustic wave flux at the photospheric boundary was varied
over several orders of magnitude.

\begin{figure}
\epsscale{1.17}
\plotone{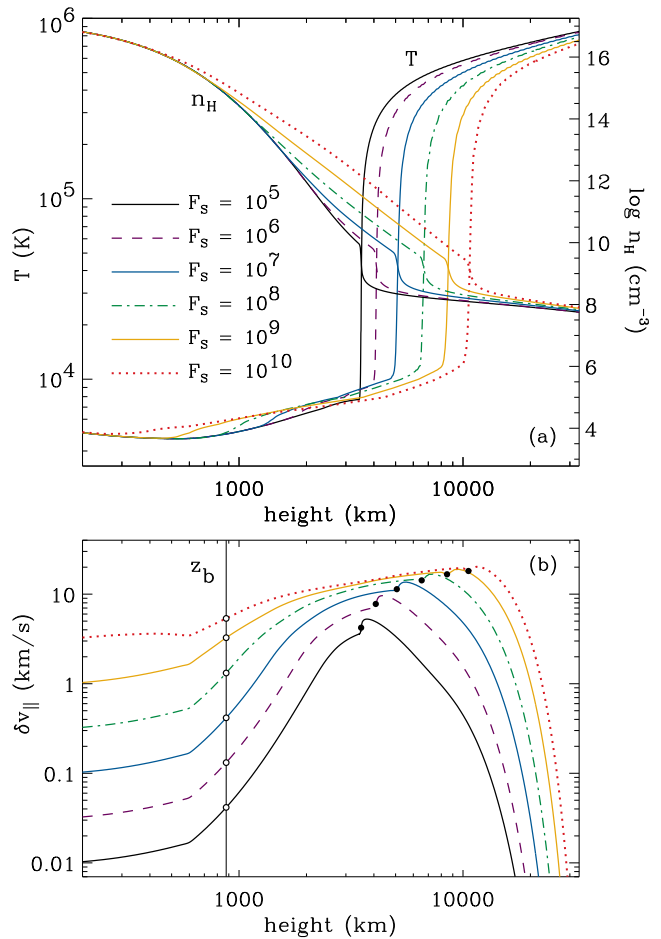}
\caption{Height dependence of (a) time-steady temperatures $T$ and
hydrogen number densities $n_{\rm H}$, and (b) acoustic wave
velocity amplitudes $\delta v_{\parallel}$, corresponding to a set of
ZEPHYR models with a range of acoustic wave fluxes $F_{\rm S}$.
Line colors/styles denote $F_{\rm S}$ (see caption) and are
consistent in both panels.
Also shown are the chromospheric base height $z_b$ (white circles)
and transition region height $z_{\rm TR}$ (black circles).
\label{fig02}}
\end{figure}

Figure~\ref{fig02} summarizes the results from the ZEPHYR models with
a range of acoustic wave power inputs.
Figure~\ref{fig02}(a) shows the time-steady variation of density and
temperature with height, and Figure~\ref{fig02}(b) shows how the
root-mean-squared (rms) parallel velocity amplitude
$\delta v_{\parallel}$ varies with height for these models.
As the acoustic waves become stronger, the chromospheric scale height
receives an increasingly large augmentation from wave pressure.
In the models of \citet{CvB07}, the sharp TR
between chromospheric and coronal temperatures occurs when the
density dips below a critical value determined by the peak of the
optically thin radiative loss function \citep[see also][]{Ow04}.
The models with stronger acoustic waves have flatter density
gradients, so the critical density is reached at larger heights.

Table~1 lists some key properties of each ZEPHYR model.
The acoustic wave energy flux $F_{\rm S}$ is injected at the
photospheric lower boundary, and the velocity amplitudes
$\delta v_{\parallel}$ increase monotonically with increasing height
from the photosphere to the TR.
Between the photosphere ($z=0$) and the TR height ($z_{\rm TR}$)
we focus on the lower chromospheric base height
$z_{b} = 876$~km and highlight it with a vertical line in
Figure~\ref{fig02}(b).
The TR is defined as the height at which the modeled temperature $T$
first rises to 20,000 K.
This is lower than the temperatures at which most TR emission lines
are formed, but we are using $z_{\rm TR}$ as an effective height for
the tips of spicules and jets; i.e., where chromospheric emission ends.

\begin{deluxetable}{cccccc}
\tablecaption{Chromospheric Properties of ZEPHYR Models
\label{table01}}
\tablewidth{0pt}
\tablehead{
\colhead{$F_{\rm S}$} &
\colhead{$\delta v_{\parallel}(z_{b})$} &
\colhead{$\delta v_{\parallel}(z_{\rm TR})$} &
\colhead{$z_{\rm TR}$} &
\colhead{$u_{1 AU}$} &
\colhead{$\dot{M}$} \\
\colhead{(erg s$^{-1}$ cm$^{-2}$)} &
\colhead{(km s$^{-1}$)} &
\colhead{(km s$^{-1}$)} &
\colhead{(km)} &
\colhead{(km s$^{-1}$)} &
\colhead{($M_{\odot}$/yr)}
}
\startdata
 0         & 0      &  0    &   3364  &  723 & $1.91 \times 10^{-14}$ \\
 $10^5$    & 0.0416 &  4.22 &   3508  &  722 & $1.91 \times 10^{-14}$ \\
 $10^6$    & 0.132  &  7.77 &   4078  &  724 & $1.88 \times 10^{-14}$ \\
 $10^7$    & 0.416  &  11.4 &   5062  &  720 & $1.90 \times 10^{-14}$ \\
 $10^8$    & 1.32   &  14.3 &   6540  &  720 & $1.88 \times 10^{-14}$ \\
 $10^9$    & 3.26   &  16.7 &   8484  &  721 & $1.86 \times 10^{-14}$ \\
 $10^{10}$ & 5.36   &  18.1 &  10576  &  728 & $1.76 \times 10^{-14}$
\enddata
\end{deluxetable}

The time-averaged shock-driven levitation of the TR is a key result
of the models shown in Figure~\ref{fig02}.
Without any compressive waves, the Sun's TR occurs in the models at
a height of $\sim$3300 km above the photosphere.
This is slightly higher than the canonical range of 2000--2500 km that
is seen in one-dimensional empirical models \citep[e.g.,][]{VAL,AL08}.
With increasing ``turbulent pressure,'' the modeled $z_{\rm TR}$ can
increase to values greater than 10,000 km.
Table~1 gives a span of heights between these extremes, and these
are roughly consistent with the values sometimes reported
from off-limb chromospheric measurements \citep{Zh98,FK00}.
In any case, we use the results given in Table~1 as an interpolation
lookup table that provides an instantaneous estimate of $z_{\rm TR}$
for any given value of $\delta v_{\parallel}(z_{b})$.

In the upper chromosphere, the acoustic waves begin to dissipate
because they steepen into shocks.
The time-averaged damping rate is dominated by the entropy change
$T \Delta S$ at each shock, spread out over the time between
successive shock passages.
The models with larger values of $F_{\rm S}$ undergo steepening
at lower heights, so their chromospheric profiles of
$\delta v_{\parallel}(z)$ become flatter and more saturated.
Above the TR, the acoustic waves damp rapidly due to the greatly
amplified rate of heat conduction in the corona
\citep[see, e.g., Equation~(26) of][]{CvB07}.
Because of this rapid damping, the acoustic waves have a negligible
effect on the eventual acceleration of the fast solar wind.\footnote{%
However, \citet{Nu13} suggested that the rapid damping of 
compressive waves may have a strong impact on the coronal heating
in large hydrostatic loops.
With enough energy dissipated at low heights, the maximum
temperature may occur below the loop apex, thus giving rise to
loops with decreasing $T(z)$ in the corona.}
Note from Table~1 that the wind speed at 1~AU and the sphere-averaged
mass loss rate $\dot{M}$ are barely affected by changing $F_{\rm S}$.

Below, we make use of the correlation seen in Table~1 between
$\delta v_{\parallel}(z_{b})$ and $z_{\rm TR}$.
This is applied to a time-dependent simulation of incompressible
turbulence that does {\em not} contain acoustic waves driven at
the photosphere.
Instead, we assume that compressive waves are generated
throughout the chromosphere by the nonlinear mode conversion
mechanism discussed in Section~\ref{sec:mode}.
We assume that the upward evolution of those compressive waves
produces a shock-driven levitation similar to that seen
in the ZEPHYR models.
Of course, it should be made clear that the second-order variations
in $\delta v_{\parallel}$ described by Equation (\ref{eq:convert})
are not identical to classical sound waves; e.g., they propagate
at a phase speed of $V_{\rm A}$ instead of $c_s$.
However, once these waves steepen into shocks, the time-averaged loss of 
momentum and energy is expected to be similar to the acoustic-wave case.

\section{Results from Time-Dependent MHD Turbulence}
\label{sec:braid}

Although the ZEPHYR code simulates the transport, cascade, and
dissipation of Alfv\'{e}nic turbulence, it does so using time-averaged
phenomenological equations.
These equations do not self-consistently simulate the actual process
of an MHD cascade, which is believed to be a consequence of partial
wave reflections and nonlinear interactions between Alfv\'{e}n wave
packets.
In order to more accurately model these processes, a time-dependent
and three-dimensional approach is needed.
We used a reduced MHD (RMHD) code called BRAID
\citep{vB11,As12,As13,vB14} that simulates the generation and evolution
of incompressible turbulence along an expanding flux tube with a
circular cross section.
This code has successfully simulated the intermittent and dynamic
heating seen in the chromospheric and coronal regions of closed loops.
Also, \citet{Sk14} proposed that this type of turbulence model may
explain the complex multi-threaded dynamics seen within Type~II spicules.

We used the simulation of an open coronal-hole flux tube developed by
\citet{WC15}.
The BRAID coronal hole model extends from the solar photosphere
($z=0$) to a maximum height of $z = 2 \, R_{\odot}$.
The choice of the latter value was a compromise between wanting to
model as much of the solar wind's acceleration region as possible
and the fact that the RMHD equations in BRAID do not yet include the
background outflow speed (i.e., they assume the wind speed
$u \ll V_{\rm A}$, which breaks down above a few solar radii).
The model was run for 2300~s of simulation time, which corresponds
to about three times the Alfv\'{e}n wave travel time through the
radial grid.

\begin{figure}
\epsscale{1.15}
\plotone{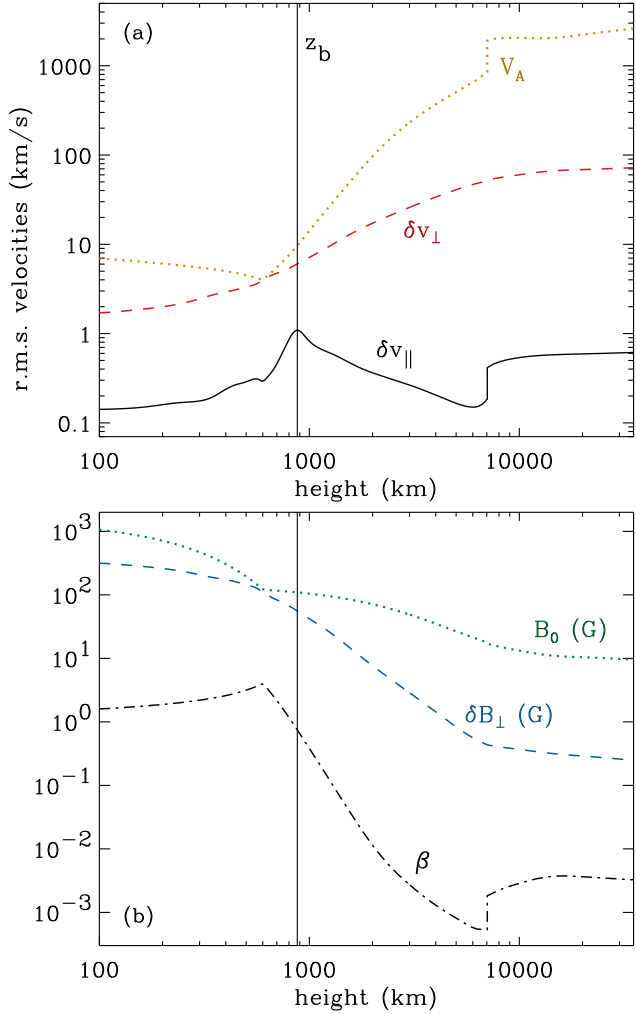}
\caption{Time-averaged plasma properties in the BRAID model of
coronal hole network:
(a) Alfv\'{e}n speed $V_{\rm A}$ (gold dotted curve), rms tranvserve
velocity amplitude $\delta v_{\perp}$ (red dashed curve),
rms longitudinal compressive wave amplitude $\delta v_{\parallel}$
(solid black curve).
(b) background magnetic field strength $B_0$ (green dotted curve),
rms transverse magnetic field amplitude $\delta B_{\perp}$
(blue dashed curve), plasma $\beta$ ratio (black dot-dashed curve).
As in Figure~\ref{fig02}(b), the base height $z_b$ is highlighted
by a vertical line.
\label{fig03}}
\end{figure}

The background properties of the BRAID model (e.g., $B_0$, $V_{\rm A}$,
$c_s$) are the same as in the time-steady polar coronal hole model 
of \citet{CvB07}.
Figure~\ref{fig03} shows a selection of these quantities from the
upper photosphere ($z = 100$~km) to the low corona.
Note that this is specifically a model of a vertical flux tube rooted
in the bright supergranular network; the field strength $B_0$ remains
greater than 100~G until one reaches a height of about 1100~km.
The magnetized network chromosphere (above the so-called ``merging
height'' where individual intergranular flux tubes join together)
can be seen in Figure~\ref{fig03} as a region of rapidly decreasing
$\beta$ with increasing height.

Figures \ref{fig03}(a) and \ref{fig03}(b) show the time-averaged
amplitudes of transverse fluctuations in velocity and magnetic field,
respectively.
As described further by \citet{WC15}, the rms averaging in these cases
was performed in two steps: (1) the variance was taken over all
92 discrete RMHD spectral wave modes at each height and time step,
then (2) the variance over simulation time was computed at each height,
excluding the earliest time steps during which the base-driven waves
had not yet traversed the grid.
Note that the BRAID fluctuations do not obey ideal MHD energy
equipartition; it is possible for the energy densities in the
kinetic and magnetic fluctuations to be unequal to one another.
It is clear that the conditions in the upper chromosphere (around
$z_b$) appear to be optimal for strong nonlinear mode conversion; not
only is $\beta \approx 1$, but also $\delta B_{\perp} \approx B_0$.

To focus on the region(s) of the chromosphere and corona in which the
mode conversion is strongest, we applied Equation (\ref{eq:convert})
to the time-averaged rms properties shown in Figure~\ref{fig03}.
The resulting height dependence of the rms $\delta v_{\parallel}$
amplitude is also plotted in Figure \ref{fig03}(a).
There is a clear peak in the low chromosphere at a height of 876~km,
with a maximum amplitude of $\sim$1 km~s$^{-1}$.
This location is defined as the base height $z_{b}$, and we consider
it as an effectively localized chromospheric ``source'' of
compressive waves.

Next we examine the highly variable and intermittent nature of the
BRAID turbulence at a fixed height.
Specifically, at the base height $z_b$, the fluctuations were
averaged over all 92 spectral wave modes (i.e., integrated over
the $k_{\perp}$ power spectrum), but not over time.
At this height, the mean value of the RMHD Alfv\'{e}n velocity
amplitude is $\delta v_{\perp} = 5.03$ km~s$^{-1}$,
with a standard deviation of 2.92 km~s$^{-1}$ and a long
tail in the distribution that extends up to a maximum
value of 13.9 km~s$^{-1}$.
For comparison, the Alfv\'{e}n speed $V_{\rm A}$ and sound speed $c_s$
at this height are 9.64 km~s$^{-1}$ and 8.35 km~s$^{-1}$, respectively.
A similarly processed time series of the ratio $\delta B_{\perp}/B_0$
has a mean of 0.45 at this height, a standard deviation
of 0.21, and a maximum value of 1.08.

\begin{figure}
\epsscale{1.13}
\plotone{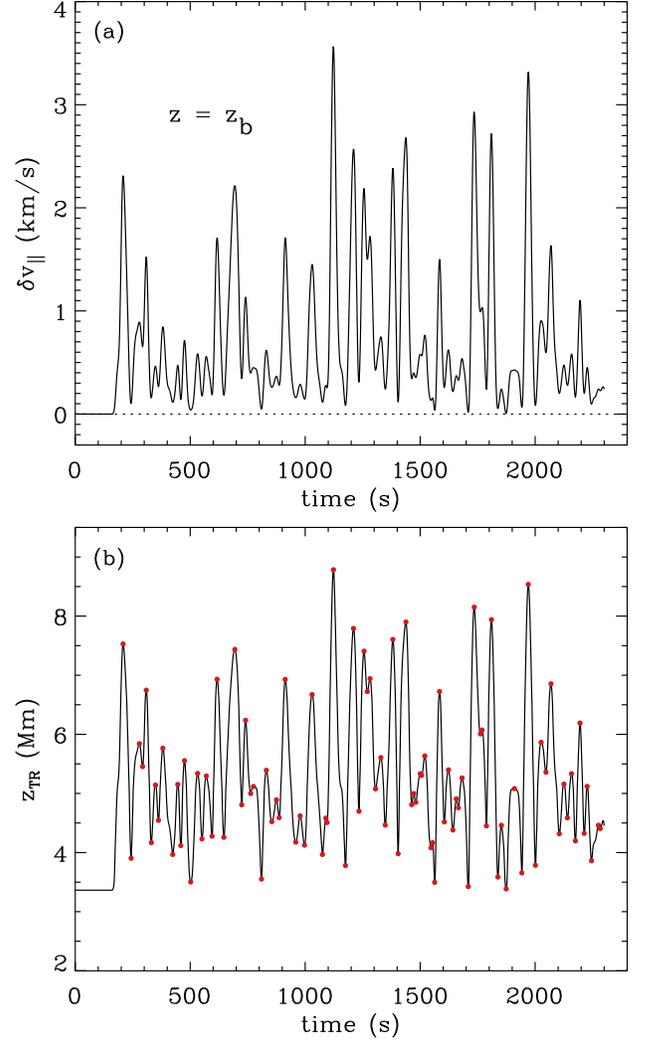}
\caption{BRAID-model time dependence of:
(a) the simulated compressive wave amplitude $\delta v_{\parallel}$
at the base height $z_b$, and (b) the instantaneous TR height
$z_{\rm TR}$ corresponding to the variable $\delta v_{\parallel}$.
Individual maxima and minima in $z_{\rm TR}$ are highlighted with
red circles.
\label{fig04}}
\end{figure}

Figure \ref{fig04}(a) shows the result of applying 
Equation (\ref{eq:convert}) to estimate the time dependence of
$\delta v_{\parallel}$ at $z_b$.
Note that this wildly fluctuating quantity is an actual {\em amplitude}
and not the full time-dependence of the parallel velocity.
If the input Alfv\'{e}n wave had been a monochromatic sinusoidal
oscillation, the estimated $\delta v_{\parallel}$ amplitude
would have been an unchanging constant.
Also, because of the nonlinear nature of the mode conversion,
$\delta v_{\parallel}(t)$ ends up having a greater relative
variability than either $\delta v_{\perp}(t)$ or $\delta B_{\perp}(t)$.
For these incompressible transverse amplitudes, the ratios of their
standard deviations to their mean values are about 0.5--0.6.
For the computed time series of $\delta v_{\parallel}$,
this ratio is about 1.0.

Figure \ref{fig04}(b) shows the time dependence of $z_{\rm TR}$,
which we computed via straightforward interpolation from the ZEPHYR
model properties given in Table~1.
The TR height varies up and down with swings of order 2 to 5 Mm,
which overlaps with the observed range of {\em IRIS} network jet
lengths \citep{Ti14}.
Implicit in Figure \ref{fig04}(b) is the assumption that the upper
chromosphere's response to variability in $\delta v_{\parallel}$
is more or less instantaneous.
A more accurate model would have to include a finite relaxation
time for the wave-pressure levitation to take effect.
Because the compressive waves travel at a phase speed $V_{\rm A}$,
we anticipate that this relaxation time should be given roughly by
the Alfv\'{e}n-wave travel time from $z_b$ to $z_{\rm TR}$.
This travel time is about 50~s, which is similar in magnitude to the
recurrence time between the modeled oscillations in $z_{\rm TR}$.
Thus, even though future simulations are needed to verify these
effects (see Section \ref{sec:conc}), we do not believe they will
differ greatly from the simpler estimates made here.

\section{Interpretation: Spicules and Jets?}
\label{sec:interp}

In the ZEPHYR models discussed in Section~\ref{sec:steep}, the
``steepened'' acoustic wave amplitudes $\delta v_{\parallel}$ at
the TR were of order 5 to 20 km~s$^{-1}$.
These velocities are small when compared to the observed velocities
of Type II spicules and {\em IRIS} network jets.
However, we see much larger {\em apparent} velocities when examining
the upward and downward variations of $z_{\rm TR}$ in
Figure~\ref{fig04}(b).
We compute these velocities by recording the minima and maxima in
$z_{\rm TR}(t)$ and taking sequential finite differences between them
in height ($\Delta z$) and in time ($\Delta t$).
These minima and maxima are shown as red symbols in Figure~\ref{fig04}(b).
The apparent velocity for each jet-like event is then computed as
$V_{\rm jet} = \Delta z / \Delta t$.
From the oscillatory nature of $z_{\rm TR}(t)$, it is apparent that
there will be roughly equal numbers of positive (low to high) and
negative (high to low) values of $V_{\rm jet}$.
We eventually plan to simulate {\em IRIS}-like emission-line images
of these evolving features, but doing so is beyond the scope of
this paper.
For now, we focus on the upward motions in $z_{\rm TR}$ (e.g., the 46
out of 93 cataloged events for which $V_{\rm jet} > 0$) and compare
them to observed upward motions in Type II spicules and network jets.

\begin{figure}
\epsscale{1.17}
\plotone{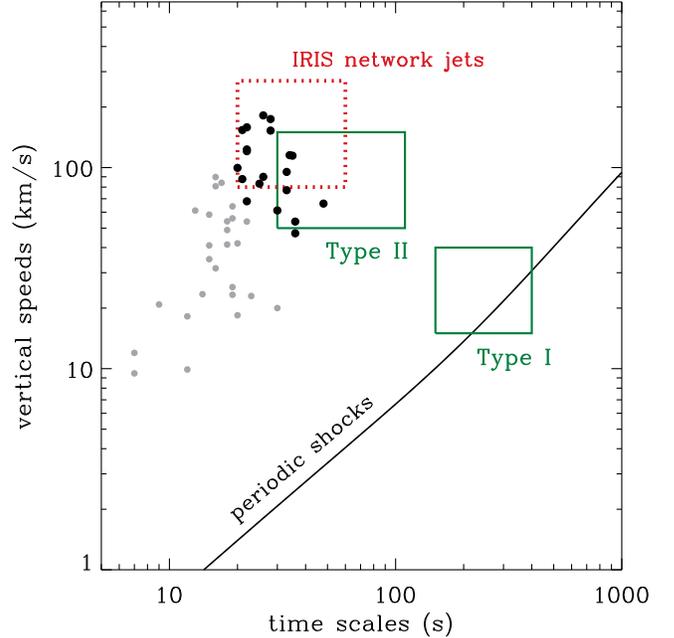}
\caption{Two-dimensional diagram of representative time scales plotted
versus vertical speeds.
Observed ranges of parameters for Type I and II spicules (green
solid-curve boxes) and {\em IRIS} network jets (red dotted-curve box)
are compared with BRAID-model simulations of $\Delta t$ and $V_{\rm jet}$.
Large black symbols show modeled events with lengths greater than 2$''$,
and small gray symbols show modeled events with lengths less than 2$''$.
The \citet{BC85} relationship for periodic shocks is shown with a
solid black curve.
\label{fig05}}
\end{figure}

Figure~\ref{fig05} shows how the modeled collection of positive
$V_{\rm jet}$ values is correlated with their ``lifetimes'' $\Delta t$.
We found in the list of 46 events that $\Delta z$ is roughly
proportional to $\Delta t^2$, so there is a roughly linear relationship
between $V_{\rm jet}$ and $\Delta t$.
In Figure~\ref{fig05} we also show reported ranges for the speeds and
lifetimes of Type I and II spicules \citep{Pe12} and {\em IRIS}
network jets \citep{Ti14,Ti15}.
The overlap between the properties of Type II spicules and network
jets has led to a growing conjecture that these represent identical
magnetic features that have been observed in different ways.
It is clear that our modeled events with the highest speeds
(corresponding also to the largest values of $\Delta z$) have very
similar properties as the observed features.

Figure~\ref{fig05} highlights modeled events with $\Delta z \geq 2''$
(i.e., $\Delta z \geq 1.45$~Mm) with larger and darker symbols.
This dividing line is close to the mean value of our distribution of
simulated jet lengths, which exhibited a range between 0.1$''$
and 7$''$.
The open BRAID flux tube has a diameter that expands with increasing
height, from about 1$''$ in the low chromosphere ($z_b$) to 3$''$ at
the largest heights shown in Figure~\ref{fig03}.
For comparison, the observed {\em IRIS} network jets have lengths
between about 3$''$ and 12$''$, and roughly constant widths of
order 0.5$''$ \citep{Ti14,Ti15}.
We believe that only events with $\Delta z$ greater than their
cross-sectional diameter would actually be observable as narrow
jet-like enhancements in intensity.
Thus, the darker symbols with $\Delta z \geq 2''$ are meant to show
only the events that would be distinctly noticeable as jets in, e.g.,
{\em IRIS} image sequences.
Shorter jets (like the modeled events shown in light gray) may exist
on the Sun, but they would likely be buried in the rapidly fluctuating
background of the underlying chromospheric network.

The solid curve in Figure~\ref{fig05} shows the relationship between
the initial upward velocities of shocks in a hydrostatic atmosphere
and the recurrence timescale between successive shocks
in a periodic train.
We solved the semi-analytic equations given in Section~II of
\citet{BC85} for a range of Mach numbers between 1.1 and 30.
In this model, gas is accelerated upward by each shock, then it
tends to fall back down after the shock passes by.
For periodic shock trains that obey the speed--timescale relation
shown by the curve, each parcel of gas ends up at the same height
one period later and executes cyclic motion.
It is interesting that the properties of Type~I spicules overlap
with this critical curve, since they often appear to show parabolic
trajectories in which parcels return to their original heights.

Regions to the left of the critical curve in Figure~\ref{fig05}
correspond to shorter timescales than are required for cyclic
motion.
Thus, a given fluid parcel would encounter the ``next'' shock before
it has fallen back to its original height.
If the shock trains completely cover the stellar surface, such rapid
recurrence would be associated with net mass loss \citep{WH79,BC85}.
Both Type~II spicules and {\em IRIS} network jets occur in this
region of parameter space, and it has been suggested that these
features feed plasma into the corona and solar wind
\citep[e.g.,][]{Dp09,Ti14}.
Chromospheric diagnostics of Type~II spicules often show bright
features moving up and not coming back down \citep{Pe12}, but
more recent coordinated observations with {\em IRIS} have shown
some parabolic-like downflows \citep{Sk15}.
Connections between observed features and the idealized periodic
shock model of \citet{BC85} are instructive and suggestive, but
they certainly do not tell the whole story.

\section{Discussion and Conclusions}
\label{sec:conc}

The goal of this paper was to explore one promising way that strong
turbulence may produce dense, short-lived, and field-aligned extensions
of chromospheric and TR plasma.
We made use of a time-dependent RMHD model of Alfv\'{e}nic
turbulence in the coronal-hole network, and we took particular notice
of the intermittent amplitude variability in the mid-chromosphere.
At a height of about 900~km above the photosphere, the wave properties
appear to be optimal to produce a spike of nonlinear mode conversion
into longitudinal, compressive fluctuations.
These waves have been shown to be able to ``puff up'' the effective
density scale height of the chromosphere and thus temporarily
increase the height of the TR.
Using an existing grid of models, we computed the time-dependent
TR height as an instantaneous response to the varying wave amplitudes
and scale heights.
Apparent upward velocities and recurrence timescales measured from the
model time series agree quite well with the observed properties of
{\em IRIS} network jets and Type~II spicules.

There are other observable characteristics of jets and spicules that
can, in principle, be compared with our models.
Do the {\em IRIS} jets appear only for specific ranges of plasma
properties in the upper chromosphere (i.e., those that maximize the
mode conversion ``spike'' at $z_b$)?
Is the observed filling factor of the jets (both in space and time)
in agreement with the intermittency seen between the largest-amplitude
pulses of $\delta v_{\parallel}$ in Figure~\ref{fig04}?
These comparisons require robust statistics from the measurement of
hundreds of individual jets and spicules.
Collecting data with sufficient accuracy may require the use of
automated feature-detection algorithms \citep[e.g.,][]{Aw13}.

We note that the models presented in this paper do not represent a
completely self-consistent simulation of the proposed jet/spicule
formation mechanism.
Instead, we attempted to show---via a sequence of separate, simple,
and well-understood models---that the various ingredients are present
at the right order of magnitude to produce the proposed effects.
We pay for this conceptual simplicity with the fact that the results
(e.g., the up/down dynamics of the TR interface) are not likely to be
quantitatively accurate.
Full three-dimensional MHD simulations are required to test these ideas.
Compressive simulations performed with fewer than three
dimensions \citep[e.g.,][]{MS14,Ko15} already show suggestive hints
of the relevant mode conversion from Alfv\'{e}n waves to
compressive/spicule-like pulses.

A comprehensive explanation for the {\em IRIS} jets and Type~II spicules
will also require taking into account some additional processes and
complexities that we did not include.  For example:
\begin{enumerate}
\item
As described in Section~\ref{sec:intro}, the larger polar plumes and
jets are probably driven by magnetic reconnection at their footpoints.
There may be an overlap between the smallest of these jets and the ones
observed by {\em IRIS} in the (mostly) unipolar network
\citep[see also][]{Mo11}.
Coronal reconnection can also generate Alfv\'{e}n waves
\citep{Ho06,Ly14}, and turbulence can also lead to the formation of
small scale reconnecting current sheets \citep[e.g.,][]{Mt15}.
The traditional dichotomy between wave-driven and
reconnection-driven coronal heating theories is no longer so
sharp, and it is useful to keep both kinds of processes in mind.
\item
Our models assumed that open magnetic flux tubes in the coronal-hole
network are essentially isolated from one another.
However, there has been a great deal of work to study how wave-like
fluctuations can enable the sharing of energy between neighboring
flux tubes and their weak-field surroundings
\citep[e.g.,][]{UK74,Rb00,Bg02,HvB08,Of09,Mm15}.
This is another type of ``mode coupling'' that needs to be
taken into account.
\item
We ignored parametric instabilities and nonlinear mode coupling that
is enhanced by inhomogeneous background properties (see
Section~\ref{sec:mode}).
The fact that we obtained the correct order-of-magnitude effect for
the jet and Type~II spicule properties may suggest that the adopted
second-order ponderomotive coupling mechanism is dominant.
However, the other proposed effects are likely to produce lower-frequency
compressive waves that are the ones observed to survive to larger heights
\citep[see, e.g.,][]{Li15}.
\end{enumerate}
Lastly, we note that the small jets and spicules discussed in this
paper may be relevant to the larger problems of coronal heating
and solar wind acceleration.
There is evidence that the rapid upward mass transfer in Type~II
spicules continues as the plasma heats to temperatures in excess of
$10^5$--$10^6$~K \citep{MD09,Sk15}.
However, there is also skepticism concerning the suggestion that
jets and spicules act as a primary source of coronal plasma
\citep{Kl12,Kl15,Ju12}.
Nevertheless, it appears more certain that the {\em waves} originating
in lower atmospheric structures survive as they propagate up into the
corona and heliosphere.
Similar kinds of nonlinear mode coupling have been proposed to act
along open field lines in the solar wind
\citep[e.g.,][]{DZ01,Ch05,CvB12,My14},
and what we learn about this process in the chromosphere and TR can
help improve our understanding of these other regions as well.

\acknowledgments

The authors gratefully acknowledge
Adriaan van Ballegooijen for developing the BRAID code and for
decades of valuable collaboration.
We also thank Hui Tian, Sean McKillop, Rebecca Arbacher, and the
{\em IRIS} team for their intrepid efforts to identify and
characterize the faint, short-lived network jets.
This work was supported by NSF SHINE program grant AGS-1540094,
NSF Graduate Research Fellowship grant DGE-1144152, and
start-up funds from the Department of Astrophysical and Planetary
Sciences at the University of Colorado Boulder.


\begin{thebibliography}{}

\bibitem[Ahmad \& Withbroe(1977)]{AW77}
Ahmad, I. A., \& Withbroe, G. L. 1977, \solphys, 53, 397

\bibitem[Aschwanden et al.(2013)]{Aw13}
Aschwanden, M. J., De Pontieu, B., \& Katrukha, E. 2013, Entropy, 15, 3007

\bibitem[Asgari-Targhi \& van Ballegooijen(2012)]{As12}
Asgari-Targhi, M., \& van Ballegooijen, A. A. 2012, \apj, 746, 81

\bibitem[Asgari-Targhi et al.(2013)]{As13}
Asgari-Targhi, M., van Ballegooijen, A. A., Cranmer, S. R., et al. 2013,
\apj, 773, 111

\bibitem[Avrett \& Loeser(2008)]{AL08}
Avrett, E. H., \& Loeser, R. 2008, \apjs, 175, 229

\bibitem[Banerjee et al.(1998)]{Bj98}
Banerjee, D., Teriaca, L., Doyle, J. G., et al. 1998, \aap, 339, 208

\bibitem[Beckers(1972)]{Be72}
Beckers, J. M. 1972, \araa, 10, 73

\bibitem[Bertschinger \& Chevalier(1985)]{BC85}
Bertschinger, E., \& Chevalier, R. A. 1985, \apj, 299, 167

\bibitem[Bogdan et al.(2002)]{Bg02}
Bogdan, T. J., Rosenthal, C. S., Carlsson, M., et al. 2002,
Astron.\  Nachr., 323, 196

\bibitem[Chandran(2005)]{Ch05}
Chandran, B. D. G. 2005, \prl, 95, 265004

\bibitem[Cheung et al.(2015)]{Cg15}
Cheung, M. C. M., De Pontieu, B., Tarbell, T. D., et al. 2015,
\apj, 801, 83

\bibitem[Cranmer \& van Ballegooijen(2012)]{CvB12}
Cranmer, S. R., \& van Ballegooijen, A. A. 2012, \apj, 754, 92

\bibitem[Cranmer et al.(2007)]{CvB07}
Cranmer, S. R., van Ballegooijen, A. A., \& Edgar, R. J. 2007,
\apjs, 171, 520

\bibitem[Culhane et al.(2007)]{Cu07}
Culhane, L., Harra, L. K., Baker, D., et al. 2007, \pasj, 59, S751

\bibitem[Del Zanna et al.(2015)]{DZ15}
Del Zanna, L., Matteini, L., Landi, S., et al. 2015,
J.\  Plasma Phys., 81, 3202

\bibitem[Del Zanna et al.(2001)]{DZ01}
Del Zanna, L., Velli, M., \& Londrillo, P. 2001, \aap, 367, 705

\bibitem[De Pontieu(1999)]{Dp99}
De Pontieu, B. 1999, \aap, 347, 696

\bibitem[De Pontieu et al.(2007)]{Dp07}
De Pontieu, B., McIntosh, S. W., Hansteen, V. H., et al. 2007,
\pasj, 59, S655

\bibitem[De Pontieu et al.(2009)]{Dp09}
De Pontieu, B., McIntosh, S. W., Hansteen, V. H., et al. 2009,
\apjl, 701, L1

\bibitem[De Pontieu et al.(2014b)]{Dp14b}
De Pontieu, B., Rouppe van der Voort, L., McIntosh, S. W., et al. 2014b,
Science, 346, 1255732

\bibitem[De Pontieu et al.(2014a)]{Dp14a}
De Pontieu, B., Title, A. M., Lemen, J. R., et al. 2014a,
\solphys, 289, 2733

\bibitem[Dewar(1970)]{Dw70}
Dewar, R. L. 1970, Phys.\  Fluids, 13, 2710

\bibitem[Dobrzycka et al.(2002)]{Do02}
Dobrzycka, D., Cranmer, S. R., Raymond, J. C., et al. 2002,
\apj, 565, 621

\bibitem[Feldman et al.(1979)]{Fe79}
Feldman, U., Doschek, G. A., \& Mariska, J. T. 1979, \apj, 229, 369

\bibitem[Filippov \& Koutchmy(2000)]{FK00}
Filippov, B., \& Koutchmy, S. 2000, \solphys, 196, 311

\bibitem[Fletcher et al.(2015)]{Fl15}
Fletcher, L., Cargill, P. J., Antiochos, S. K., \& Gudiksen, B. V.
2015, \ssr,188, 211

\bibitem[Goldstein(1978)]{Go78}
Goldstein, M. L. 1978, \apj, 219, 700

\bibitem[Hasan \& van Ballegooijen(2008)]{HvB08}
Hasan, S. S., \& van Ballegooijen, A. A. 2008, \apj, 680, 1542

\bibitem[Heyvaerts \& Priest(1983)]{HP83}
Heyvaerts, J., \& Priest, E. R. 1983, \aap, 117, 220

\bibitem[Hollweg(1971)]{Ho71}
Hollweg, J. V. 1971, \jgr, 76, 5155

\bibitem[Hollweg(1992)]{Ho92}
Hollweg, J. V. 1992, \apj, 389, 731

\bibitem[Hollweg(2006)]{Ho06}
Hollweg, J. V. 2006, Phil.\  Trans.\  Roy.\  Soc.\  A, 364, 505

\bibitem[Hollweg et al.(1982)]{HJG}
Hollweg, J. V., Jackson, S., \& Galloway, D. 1982, \solphys, 75, 35

\bibitem[Hollweg \& Kaghashvili(2012)]{HK12}
Hollweg, J. V., \& Kaghashvili, E. K. 2012, \apj, 744, 114

\bibitem[Jacques(1977)]{J77}
Jacques, S. A. 1977, \apj, 215, 942

\bibitem[Jayanti \& Hollweg(1993)]{JH93}
Jayanti, V., \& Hollweg, J. V. 1993, \jgr, 98, 19049

\bibitem[Jess et al.(2009)]{Js09}
Jess, D. B., Mathioudakis, M., Erd\'{e}lyi, R., et al. 2009,
Science, 323, 1582

\bibitem[Judge et al.(2012)]{Ju12}
Judge, P. G., De Pontieu, B., McIntosh, S. W., et al. 2012,
\apj, 746, 158

\bibitem[Klimchuk(2012)]{Kl12}
Klimchuk, J. A. 2012, \jgr, 117, A12102

\bibitem[Klimchuk(2015)]{Kl15}
Klimchuk, J. A. 2015, Phil.\  Trans.\  Roy.\  Soc.\  A, 373, 20140256

\bibitem[Kono et al.(2015)]{Ko15}
Kono, S., Yokoyama, T., Toriumi, S., \& Katsukawa, Y. 2015,
IRIS-4 Workshop, poster 19

\bibitem[Krishna Prasad et al.(2012)]{KP12}
Krishna Prasad, S., Banerjee, D., Van Doorsselaere, T., \&
Singh, J. 2012, \aap, 546, A50

\bibitem[Kudoh \& Shibata(1999)]{KS99}
Kudoh, T., \& Shibata, K. 1999, \apj, 514, 493

\bibitem[Kukhianidze et al.(2006)]{Ku06}
Kukhianidze, V., Zaqarashvili, T. V., \& Khutsishvili, E. 2006,
\aap, 449, L35

\bibitem[Lamb(1908)]{Lb08}
Lamb, H. 1908, Proc.\  London Math.\  Soc., 7, 122

\bibitem[Lamb(1932)]{Lb32}
Lamb, H. 1932, Hydrodynamics (New York: Dover Publications)

\bibitem[Lee \& Roberts(1986)]{LR86}
Lee, M. A., \& Roberts, B. 1986, \apj, 301, 430

\bibitem[Liu et al.(2014)]{Li14}
Liu, J., McIntosh, S. W., De Moortel, I., et al. 2014, \apj, 797, 7

\bibitem[Liu et al.(2015)]{Li15}
Liu, J., McIntosh, S. W., De Moortel, I., et al. 2015, \apj, 806, 273

\bibitem[Ludwig \& Ku\v{c}inskas(2012)]{LK12}
Ludwig, H.-G., \& Ku\v{c}inskas, A. 2012, \aap, 547, A118

\bibitem[Lynch et al.(2014)]{Ly14}
Lynch, B. J., Edmondson, J. K., \& Li, Y. 2014, \solphys, 289, 3043

\bibitem[Mariska \& Hollweg(1985)]{MH85}
Mariska, J. T., \& Hollweg, J. V. 1985, \apj, 296, 746

\bibitem[Mathioudakis et al.(2013)]{Ma13}
Mathioudakis, M., Jess, D. B., \& Erd\'{e}lyi, R. 2013, \ssr, 175, 1

\bibitem[Matsumoto \& Shibata(2010)]{MS10}
Matsumoto, T., \& Shibata, K. 2010, \apj, 710, 1857

\bibitem[Matsumoto \& Suzuki(2014)]{MS14}
Matsumoto, T., \& Suzuki, T. K. 2014, \mnras, 440, 971

\bibitem[Matthaeus et al.(2015)]{Mt15}
Matthaeus, W. H., Wan, M., Servidio, S., et al. 2015,
Phil.\  Trans.\  Roy.\  Soc.\  A, 373, 20140154

\bibitem[McIntosh(2012)]{Mc12}
McIntosh, S. W. 2012, \ssr, 172, 69

\bibitem[McIntosh \& De Pontieu(2009)]{MD09}
McIntosh, S. W., \& De Pontieu, B. 2009, \apjl, 706, L80

\bibitem[Medvedev \& Diamond(1996)]{MD96}
Medvedev, M. V., \& Diamond, P. H. 1996, Phys.\  Plasmas, 3, 863

\bibitem[Miyamoto et al.(2014)]{My14}
Miyamoto, M., Imamura, T., Tokumaru, M., et al. 2014,
\apj, 797, 51

\bibitem[Mjolhus \& Wyller(1986)]{MW86}
Mjolhus, E., \& Wyller, J. 1986, Physica Scripta, 33, 442

\bibitem[Moore et al.(2010)]{Mo10}
Moore, R. L., Cirtain, J. W., Sterling, A. C., et al. 2010, \apj,
720, 757

\bibitem[Moore et al.(2011)]{Mo11}
Moore, R. L., Sterling, A. C., Cirtain, J. W., et al. 2011, \apjl,
731, L18
 
\bibitem[Moriyasu et al.(2004)]{Mo04}
Moriyasu, S., Kudoh, T., Yokoyama, T., et al. 2004, \apjl, 601, L107

\bibitem[Mumford et al.(2015)]{Mm15}
Mumford, S. J., Fedun, V., \& Erd\'{e}lyi, R. 2015, \apj, 799, 6

\bibitem[Murawski et al.(2015)]{Mu15}
Murawski, K., Solov'ev, A., Musielak, Z. E., et al. 2015,
\aap, 577, A126

\bibitem[Nakariakov et al.(1998)]{Nk98}
Nakariakov, V. M., Roberts, B., \& Murawski, K. 1998, \aap, 332, 795

\bibitem[Newkirk \& Harvey(1968)]{NH68}
Newkirk, G., Jr., \& Harvey, J. 1968, \solphys, 3, 321

\bibitem[Nuevo et al.(2013)]{Nu13}
Nuevo, F. A., Huang, Z., Frazin, R., et al. 2013, \apj, 773, 9

\bibitem[Ofman(2009)]{Of09}
Ofman, L. 2009, \ssr, 149, 153

\bibitem[Ofman et al.(1999)]{Of99}
Ofman, L., Nakariakov, V. M., \& DeForest, C. E. 1999, \apj, 514, 441

\bibitem[Owocki(2004)]{Ow04}
Owocki, S. P. 2004, in EAS Publ.\  Ser., 13, 163

\bibitem[Pant et al.(2015)]{Pa15}
Pant, V., Dolla, L., Muzumder, R., et al. 2015, \apj, 807, 71

\bibitem[Paraschiv et al.(2015)]{Pv15}
Paraschiv, A. R., Bemporad, A., \& Sterling, A. C. 2015, \aap, 579, A96

\bibitem[Pariat et al.(2009)]{Pa09}
Pariat, E., Antiochos, S. K., \& DeVore, C. R. 2009, \apj, 691, 61

\bibitem[Pereira et al.(2012)]{Pe12}
Pereira, T. M. D., De Pontieu, B., \& Carlsson, M. 2012, \apj, 759, 18

\bibitem[Perez \& Chandran(2013)]{PC13}
Perez, J. C., \& Chandran, B. D. G. 2013, \apj, 776, 124

\bibitem[Peter(2013)]{Pe13}
Peter, H. 2013, \solphys, 288, 531

\bibitem[Raouafi et al.(2008)]{Ra08}
Raouafi, N-.E., Petrie, G. J. D., Norton, A. A., et al. 2008,
\apjl, 682, L137

\bibitem[Roberts(2000)]{Rb00}
Roberts, B. 2000, \solphys, 193, 139

\bibitem[Shibata et al.(2007)]{Sh07}
Shibata, K., Nakamura, T., Matsumoto, T., et al. 2007, Science,
318, 1591

\bibitem[Skogsrud et al.(2014)]{Sk14}
Skogsrud, H., Rouppe van der Voort, L., \& De Pontieu, B. 2014,
\apjl, 795, L23

\bibitem[Skogsrud et al.(2015)]{Sk15}
Skogsrud, H., Rouppe van der Voort, L., De Pontieu, B., \&
Pereira, T. M. D. 2015, \apj, 806, 170

\bibitem[Spangler(1989)]{Sp89}
Spangler, S. R. 1989, Phys.\  Fluids B, 1, 1738

\bibitem[Stangalini et al.(2015)]{St15}
Stangalini, M., Giannattasio, F., \& Jafarzadeh, S. 2015, \aap, 577, A17

\bibitem[Stein \& Schwartz(1972)]{SS72}
Stein, R. F., \& Schwartz, R. A. 1972, \apj, 177, 807

\bibitem[Sterling(2000)]{St00}
Sterling, A. C. 2000, \solphys, 196, 79

\bibitem[Tavabi et al.(2015)]{Tv15}
Tavabi, E., Koutchmy, S., Ajabshirizadeh, A., et al. 2015,
\aap, 573, A4

\bibitem[Threlfall et al.(2013)]{Th13}
Threlfall, J., De Moortel, I., McIntosh, S. W., et al. 2013,
\aap, 556, A124

\bibitem[Thurgood \& McLaughlin(2013)]{TM13}
Thurgood, J. O., \& McLaughlin, J. A. 2013, \solphys, 288, 205

\bibitem[Tian et al.(2014)]{Ti14}
Tian, H., DeLuca, E., Cranmer, S. R., et al. 2014, Science, 346, 1255711

\bibitem[Tian et al.(2015)]{Ti15}
Tian, H., DeLuca, E., Cranmer, S. R., et al. 2015, IRIS-4 Workshop,
poster 38

\bibitem[Tomczyk \& McIntosh(2009)]{TM09}
Tomczyk, S., \& McIntosh, S. W. 2009, \apj, 697, 1384

\bibitem[Tomczyk et al.(2007)]{To07}
Tomczyk, S., McIntosh, S. W., Keil, S. L., et al. 2007, Science,
317, 1192

\bibitem[Tsiropoula et al.(2012)]{Ts12}
Tsiropoula, G., Tziotziou, K., Kontogiannis, I., et al. 2012,
\ssr, 169, 181

\bibitem[Turkmani \& Torkelsson(2003)]{TT03}
Turkmani, R., \& Torkelsson, U. 2003, \aap, 409, 813

\bibitem[Uchida \& Kaburaki(1974)]{UK74}
Uchida, Y., \& Kaburaki, O. 1974, \solphys, 35, 451

\bibitem[van Ballegooijen et al.(2014)]{vB14}
van Ballegooijen, A. A., Asgari-Targhi, M., \& Berger, M. A.
2014, \apj, 787, 87

\bibitem[van Ballegooijen et al.(2011)]{vB11}
van Ballegooijen, A. A., Asgari-Targhi, M., Cranmer, S. R., et al.
2011, \apj, 736, 3

\bibitem[Vasquez \& Hollweg(1996)]{VH96}
Vasquez, B. J., \& Hollweg, J. V. 1996, \jgr, 101, 13527

\bibitem[Velli \& Liewer(1999)]{VL99}
Velli, M., \& Liewer, P. 1999, \ssr, 87, 339

\bibitem[Vernazza et al.(1981)]{VAL}
Vernazza, J. E., Avrett, E. H., \& Loeser, R. 1981, \apjs, 45, 635

\bibitem[Wang \& Sheeley(1995)]{WS95}
Wang, Y.-M., \& Sheeley, N. R., Jr. 1995, \apj, 452, 457

\bibitem[Wang et al.(1998)]{Wa98}
Wang, Y.-M., Sheeley, N. R., Jr., Socker, D. G., et al. 1998,
\apj, 508, 899

\bibitem[Willson \& Hill(1979)]{WH79}
Willson, L. A., \& Hill, S. J. 1979, \apj, 228, 854

\bibitem[Woolsey \& Cranmer(2014)]{WC14}
Woolsey, L. N., \& Cranmer, S. R. 2014, \apj, 787, 160

\bibitem[Woolsey \& Cranmer(2015)]{WC15}
Woolsey, L. N., \& Cranmer, S. R. 2015, \apj, in press, arXiv:1509.00377

\bibitem[Young \& Muglach(2014)]{YM14}
Young, P. R., \& Muglach, K. 2014, \pasj, 66, S129

\bibitem[Zhang et al.(1998)]{Zh98}
Zhang, J., White, S. M., \& Kundu, M. R. 1998, \apjl, 504, L127

\bibitem[Zhdankin et al.(2015)]{Zd15}
Zhdankin, V., Uzdensky, D. A., \& Boldyrev, S. 2015, \prl, 114, 065002

\end{thebibliography}
\end{document}